\documentclass[aps,prl,numerical,linenumbers,superscriptaddress,showpacs,amsfonts,floatfix,reprint]{revtex4-1}

\usepackage{bm}
\usepackage{graphicx}
\usepackage{amsmath}

\def\la{\langle}
\def\ra{\rangle}

\def\snn{\mbox{$\sqrt{s_{_{\rm NN}}}$}}

\def\pt{p_{\rm{T}}}

\def\psin{\Psi_{\rm n}}
\def\cent{{\rm cent}}

\newcommand{\be}{\begin{eqnarray}}
\newcommand{\ee}{\end{eqnarray}}

\def\v2{\mbox{$v_2$}}

\usepackage{ulem}
\usepackage{color}

\begin{document}


\title{Acoustic scaling of anisotropic flow in shape-engineered events: implications for \\ extraction 
of the specific shear viscosity of the quark gluon plasma}

\author{ Roy~A.~Lacey}
\email[E-mail: ]{Roy.Lacey@Stonybrook.edu}
\affiliation{Department of Chemistry, 
Stony Brook University, \\
Stony Brook, NY, 11794-3400, USA}
\affiliation{Department of Physics and Astronomy, Stony Brook University, \\
Stony Brook, NY, 11794-3800}
\author{D. Reynolds} 
\affiliation{Department of Chemistry, 
Stony Brook University, \\
Stony Brook, NY, 11794-3400, USA}
\author{A.~Taranenko}
\affiliation{Department of Chemistry, 
Stony Brook University, \\
Stony Brook, NY, 11794-3400, USA} 
\author{ N.~N.~Ajitanand} 
\affiliation{Department of Chemistry, 
Stony Brook University, \\
Stony Brook, NY, 11794-3400, USA}
\author{ J.~M.~Alexander}
\affiliation{Department of Chemistry, 
Stony Brook University, \\
Stony Brook, NY, 11794-3400, USA}
\author{ Fu-Hu Liu}
\affiliation{Department of Chemistry, 
Stony Brook University, \\
Stony Brook, NY, 11794-3400, USA}
\affiliation{Institute of Theoretical Physics, Shanxi University, \\ 
Taiyuan, Shanxi 030006, China
}
\author{Yi Gu} 
\affiliation{Department of Chemistry, 
Stony Brook University, \\
Stony Brook, NY, 11794-3400, USA}
%
%
%
\author{A. Mwai}
\affiliation{Department of Chemistry, 
Stony Brook University, \\
Stony Brook, NY, 11794-3400, USA}
%
%



\date{\today}

\begin{abstract}
It is shown that the acoustic scaling patterns of anisotropic flow for different event shapes at a fixed 
collision centrality (shape-engineered events), provide robust constraints for the event-by-event 
fluctuations in the initial-state density distribution from ultrarelativistic heavy ion collisions.
The empirical scaling parameters also provide a dual-path method for extracting the specific shear viscosity 
$(\eta/s)_\mathrm{QGP}$ of the quark-gluon plasma (QGP) produced in these collisions. 
A calibration of these scaling parameters via detailed viscous hydrodynamical model calculations, gives $(\eta/s)_\mathrm{QGP}$ 
estimates for the plasma produced in collisions of Au+Au ($\sqrt{s_{NN}}= 0.2$ TeV) and Pb+Pb ($\sqrt{s_{NN}}= 2.76$ TeV). 
The estimates are insensitive to the initial-state geometry models considered.

\end{abstract}


\date{\today}

\maketitle


Considerable attention has been given to the study of anisotropic flow measurements
in heavy-ion collisions at both the Relativistic Heavy Ion Collider (RHIC) and the Large Hadron Collider (LHC)
\cite{Teaney:2003kp,Lacey:2006pn,Lacey:2006bc,*Adare:2006nq,*Drescher:2007cd,*Dusling:2007gi,%
*Xu:2007jv,*Molnar:2008xj,*Lacey:2009xx,*Dusling:2009df,*Chaudhuri:2009hj,*Lacey:2010fe,%
Romatschke:2007mq,Luzum:2008cw,*Luzum:2009sb,Song:2010mg,Aamodt:2010pa,*Luzum:2010ag,Lacey:2010ej,%
Bozek:2010wt,Hirano:2010jg,Schenke:2010rr,*Schenke:2011tv,Song:2011qa,*Shen:2010uy,*Gardim:2012yp,%
Niemi:2012ry,Qin:2010pf}. Recently, the attack has focused on studies of initial state fluctuations and 
their role in the extraction of the specific shear viscosity (i.e. the ratio of shear viscosity 
to entropy density $\eta/s$) of the quark-gluon plasma (QGP) .
These flow measurements are routinely quantified as a function of collision centrality (cent) 
and particle transverse momentum $\pt$ by the Fourier coefficients $v_n$
\begin{equation}
 v_n(\pt,\cent) = \la \cos[n(\phi-\psin)] \ra.
\end{equation}
Here $\phi$ is the azimuthal angle of an emitted particle and $\psin$ is the estimated azimuth of 
the $n$-th order event plane \cite{Ollitrault:1992bk,Adare:2010ux}; brackets denote averaging over particles and events.
The current measurements for charged hadrons~\cite{Lacey:2011av,Adare:2011tg,*ALICE:2011ab,*ATLAS:2012at,%
*Sorensen:2011fb,*Sanders:2012iz} indicate significant odd and even $v_n$ 
coefficients up to about the sixth harmonic.

The estimates of $(\eta/s)_\mathrm{QGP}$ from these $v_n$ measurements have indicated a small 
value (i.e.  1-3 times the lower conjectured bound of ${1}/{4\pi}$ \cite{Kovtun:2004de}).
Substantial theoretical uncertainties have been assigned primarily to incomplete knowledge 
of the initial-state geometry and its associated event-by-event fluctuations. 
Indeed, an uncertainty of ${\cal O}(100\%)$ in the value 
of $(\eta/s)_\mathrm{QGP}$ extracted from $v_2$ measurements at RHIC ($\snn =0.2$~TeV) \cite{Luzum:2008cw,Song:2010mg}, 
has been attributed to a {$\sim 20$\%} uncertainty in the theoretical estimates \cite{Hirano:2005xf,Drescher:2006pi,
*Drescher:2007ax,*Qiu:2011iv}
for the event-averaged initial eccentricity $\varepsilon_{2}$ of the collision zone. Here, it is important to
note that a robust method of extraction should not depend on the initial geometrical conditions since  
$(\eta/s)_\mathrm{QGP}$ is only a property of the medium itself.

Recent attempts to reduce the uncertainty for $(\eta/s)_\mathrm{QGP}$ have 
focused on: (i) the development of a more constrained 
description of the fluctuating initial-state geometry \cite{Schenke:2012wb}, (ii) the combined analysis of $v_2$ and 
$v_3$ \cite{Lacey:2010hw,Adare:2011tg,*ALICE:2011yba,Shen:2011zc,*Qiu:2011fi} and other higher order 
harmonics \cite{Schenke:2010rr,Schenke:2011tv} and (iii) a search for new constraints via ``acoustic scaling'' 
of $v_n$ \cite{Lacey:2013is,Lacey:2011ug,Lacey:2013qua}. 
The latter two approaches [(ii) and (iii)] utilize the prediction that the strength of the dissipative effects 
which influence the magnitude of $v_{n}(\cent)$,
grow exponentially as $n^2$ and $1/\bar{R}$ \cite{Lacey:2013is,Shuryak:2013ke,Lacey:2011ug}; 
\be
\frac{v_n(\text{cent})}{\varepsilon_n(\text{cent})} \propto \exp{\left(-\beta \frac{n^2}{{\bar{R}}} \right)},
\,\,\, \beta \sim \frac{4}{3} \frac{\eta}{Ts},
\label{eq:2}
\ee 
where $\varepsilon_n$ is the $n$-th order eccentricity moment, $T$ is the temperature and $\bar{R}$ is 
the initial-state transverse size of the collision zone.
Thus, a characteristic linear dependence of $\ln(v_n/\varepsilon_n)$ on $n^2$ and $1/\bar{R}$ [cf. Eq.~\ref{eq:2}], 
with slopes $\beta'\propto (\eta/s)_\mathrm{QGP}$ and $\beta''\propto (\eta/s)_\mathrm{QGP}$ are to be expected.

These scaling patterns have indeed been validated and shown to give important constraints 
for the extraction of $(\eta/s)_\mathrm{QGP}$ from both RHIC ($\snn =0.2$~TeV) and LHC ($\snn =2.76$~TeV)
data \cite{Lacey:2013is,Lacey:2011ug}. 
%
\begin{figure*}
  \begin{tabular}{ccc}
  \includegraphics[width=0.32\linewidth]{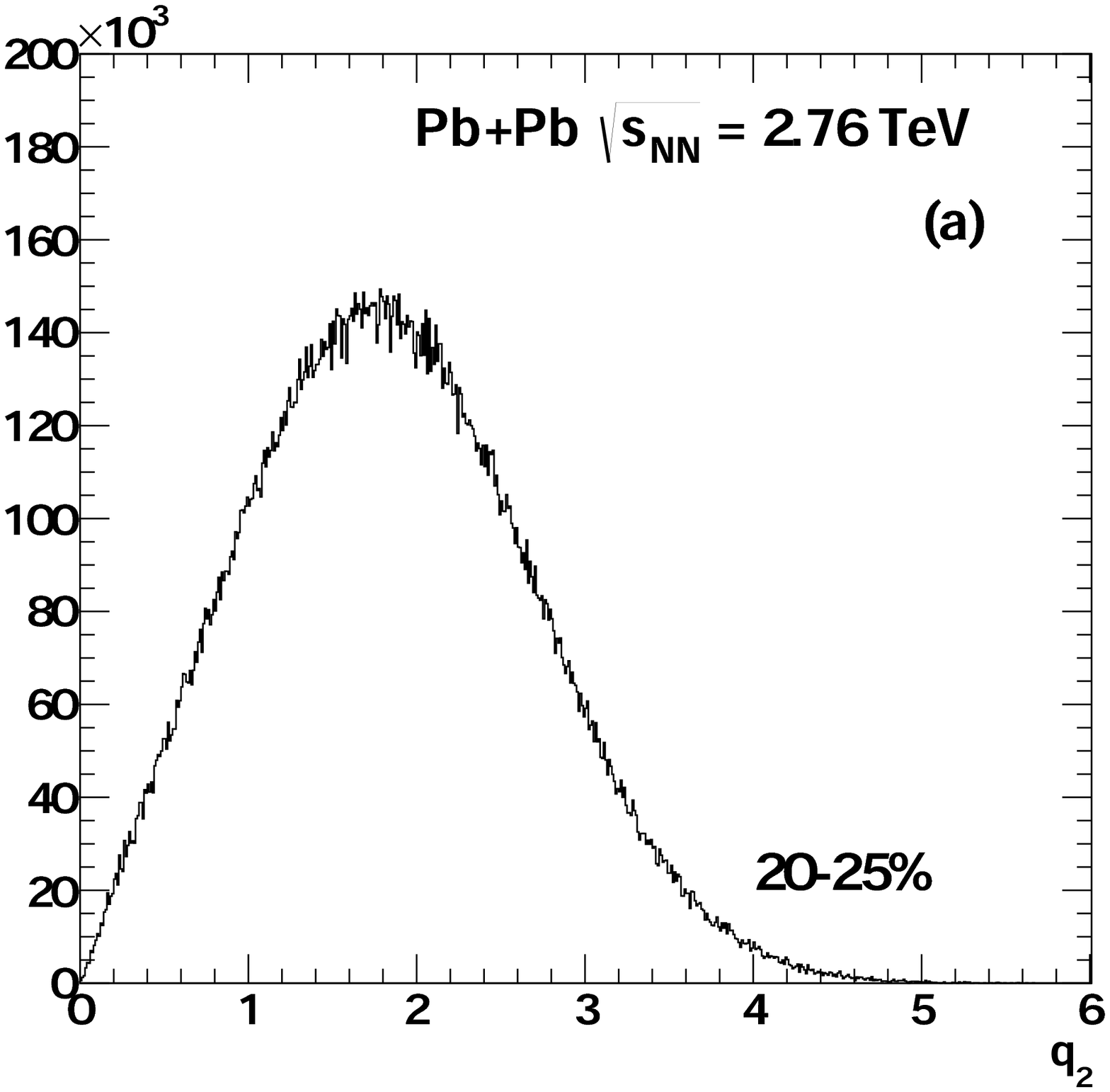} &
  \includegraphics[width=0.34\linewidth]{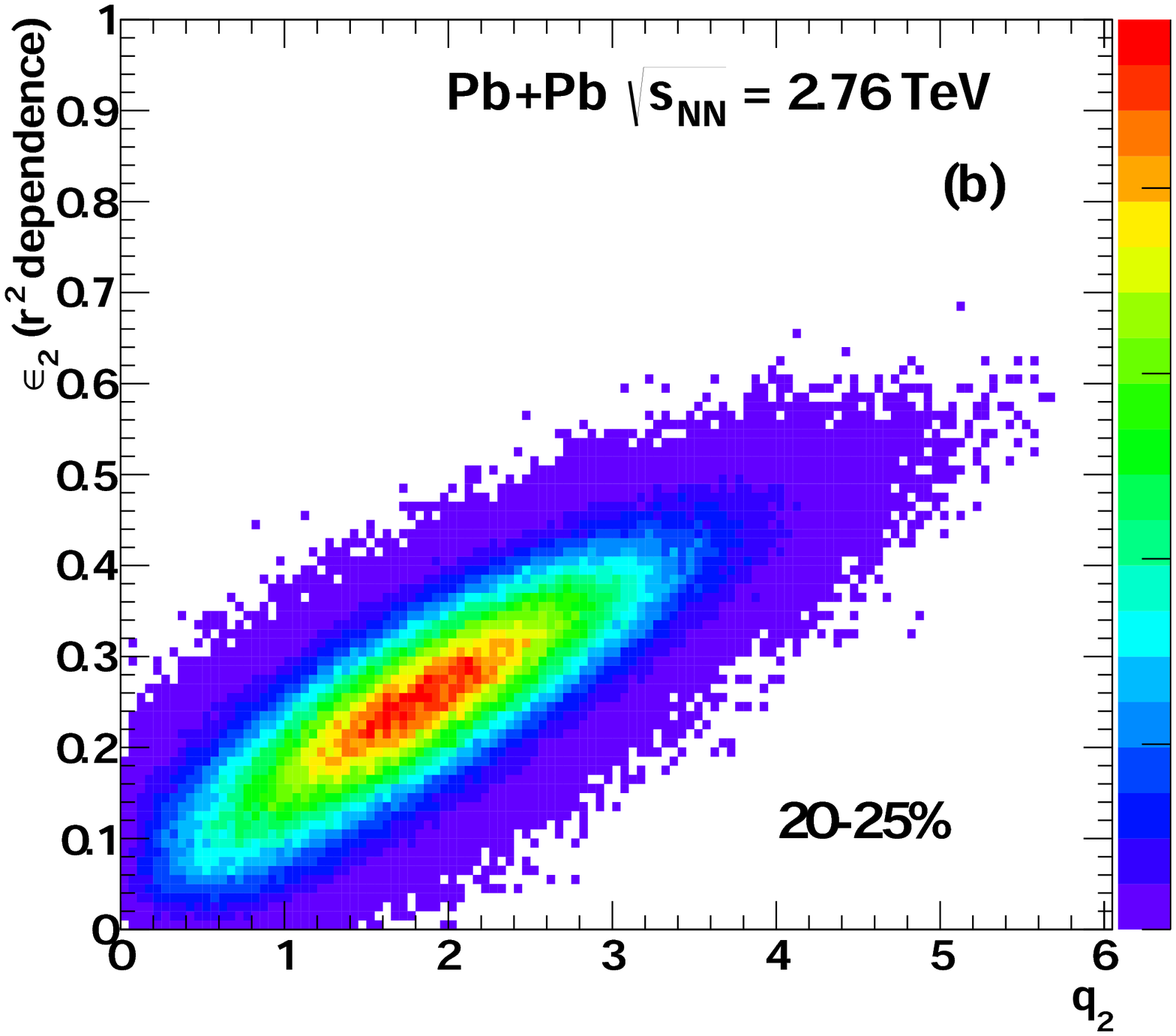} &
	\includegraphics[width=0.34\linewidth]{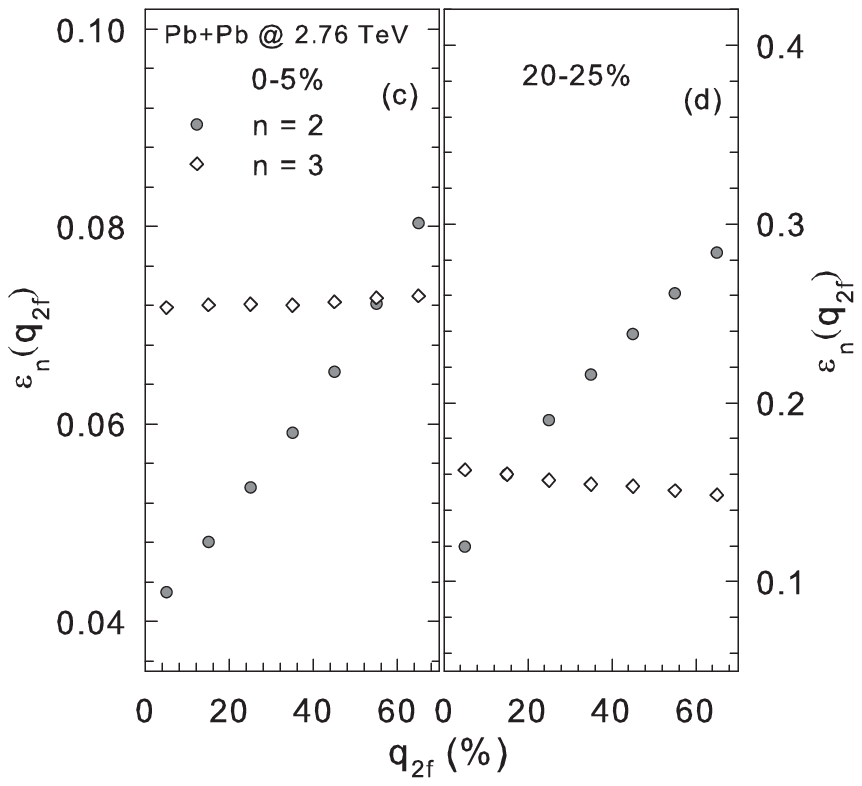}
  \end{tabular}
  \caption{(Color online) Calculated values for (a) the $q_2$ distribution for 20-25\% central events; 
	(b) $\varepsilon_2$ vs. $q_2$ for 20-25\% central events; (c) $\varepsilon_{2,3}$ vs. $q_{2\text{f}}$ 
	for 0-5\% central events; (d) $\varepsilon_{2,3}$ vs. $q_{2\text{f}}$ for 20-25\% central events. 
	The calculations were made for Pb+Pb collisions at $\snn =2.76$~TeV with the MC-Glauber model.
	}
  \label{fig:1}
\end{figure*}
%
Here, we explore a more explicit constraint for initial-state shape fluctuations, via 
scaling studies of $v_n$ measurements obtained for shape-engineered events, i.e. different event 
shapes at a fixed centrality. 

Such a constraint is derived from the expectation that the 
event-by-event fluctuations in anisotropic flow, result primarily from fluctuations in the 
size and shape (eccentricity) of the initial-state density distribution. 
Thus, various cuts on the full distribution of initial shapes 
[at a given centrality], should result in changes in the magnitudes 
of $\la\varepsilon_n\ra$, $\la\bar{R_n}\ra$ and $\la v_n\ra$. Note however, that acceptable models for 
the initial-state fluctuations should give $\la\varepsilon_n\ra$ and $\la\bar{R_n}\ra$ values 
each of which lead to acoustic scaling of $\la v_n\ra$ with little, if any, change in the 
slope parameter $\beta'$ ($\beta''$) for different event shape selections, 
i.e., $\beta'$ ($\beta''$)~$\propto (\eta/s)_\mathrm{QGP}$ is a property of the medium, not
the initial state geometry.

The $q_n$ flow vector has been proposed \cite{Schukraft:2012ah} as a tool 
to select different initial shapes from the distribution of initial-state geometries 
at a fixed centrality;
\be
&&Q_{n,x}=\sum_i^M \cos(n\phi_i);\;   
Q_{n,y}=\sum_i^M \sin(n\phi_i);\;
\\
&&q_n=Q_n/\sqrt{M},  \;
\label{eq:q} 
\ee
where $M$ is the particle multiplicity and $\phi_i$ are the azimuthal angles of the 
particles in the sub-event used to determine $q_n$.   
We use this technique for model-based evaluations of $\varepsilon_2(q_2,\cent)$ and $\bar{R}(q_2,\cent)$ 
to perform validation tests for acoustic scaling of recent $v_2(q_2,\cent)$ measurements, 
as well as to determine if $\beta''$ is independent of event shape. 
We use the acoustic scaling patterns, indicated in the results of $q_n$-averaged viscous hydrodynamical 
calculations \cite{Song:2011hk,*cms_ulc_note}, to calibrate $\beta'\text{ and }\beta''$
and make estimates of $(\eta/s)_\mathrm{QGP}$ for the plasma produced in Au+Au and Pb+Pb 
collisions at RHIC and the LHC respectively.

 

	The data employed in this work are taken from measurements by the ALICE and CMS
collaborations for Pb+Pb collisions at $\sqrt{s_{NN}}$ = 2.76 TeV \cite{Dobrin:2012zx,Chatrchyan:2012ta},
as well as measurements by the STAR collaboration for Au+Au collisions 
at $\sqrt{s_{NN}}= 200$ GeV \cite{Adams:2004bi,Aamodt:2010pa}. 
The ALICE measurements \cite{Dobrin:2012zx} exploit a three subevents technique to 
evaluate $v_2(q_2,\cent)$, where the first subevent $\text{SE}_1$ is used to 
determine $q_2$, and the particles in the second subevent $\text{SE}_2$ are used to
evaluate $v_2(q_2,\cent)$ relative to the $\Psi_{2}$ event plane determined 
from the particles in the third subevent $\text{SE}_3$. 
To suppress non-flow correlations, the detector subsystems used to select $\text{SE}_{1,2,3}$ 
were chosen so as to give a sizable pseudo-rapidity gap ($\Delta\eta_p $) between 
the particles in different subevents. 
For each centrality, $v_2(q_2)$ measurements were 
made for the full $q_2$ distribution [$v_2(q_{2(\text{Avg.})})$], as well as for events with the 
10\% lowest [$v_2(q_{2(\text{Lo})})$] and 5\% highest [$v_2(q_{2(\text{Hi})})$] values of the $q_2$ 
distribution. 

The CMS~\cite{cms_ulc_note} and STAR~\cite{Adams:2004bi} $v_n(\cent)$ measurements  
for $n = 2-6$ (CMS) and $n = 2$ (STAR) were selected to ensure compatibility with the
viscous hydrodynamical calculations discussed below. An explicit selection on $q_n$ was 
not used for these measurements; instead, they were averaged over the respective $q_n$ 
distributions to give $v_n(q_{n(\text{Avg.})},\cent) \equiv v_n(\cent)$. The systematic 
errors for the ALICE, CMS and STAR measurements are reported 
in Refs.~~\cite{Dobrin:2012zx}, \cite{Chatrchyan:2012ta} and \cite{Adams:2004bi} respectively.

Monte Carlo versions were used for (a) the Glauber (MC-Glauber)~\cite{Miller:2007ri,*Alver:2006wh} 
and (b) Kharzeev-Levin-Nardi~\cite{Kharzeev:2000ph,Lappi:2006xc,Drescher:2007ax} (MC-KLN) models for 
fluctuating initial conditions. Each was used to compute the number of participants N$_\text{part}(\text{cent})$, 
$q_n(\cent)$, $\varepsilon_n({\text{cent}})$ 
[with weight $\omega(\mathbf{r_{\perp}}) = \mathbf{r_{\perp}\!}^n$]
and $\bar{R}_n({\text{cent}})$ from the two-dimensional profile of the density of 
sources in the transverse  plane $\rho_s(\mathbf{r_{\perp}})$ \cite{Lacey:2010hw}, where 
${1}/{\bar{R}_2}~=~\sqrt{\left({1}/{\sigma_x^2}+{1}/{\sigma_y^2}\right)}$, 
with $\sigma_x$ and $\sigma_y$ the respective root-mean-square widths of the 
density distributions. Computations for these initial-state geometric quantities were also made 
for 5\% and 10\% increments in $q_n$, from the lowest ($q_{n(\text{Lo})}$) to the highest ($q_{n(\text{Hi})}$) values 
of the $q_n$ distribution. The computations were performed for both Au+Au ($\sqrt{s_{NN}}= 0.2$ TeV) 
and Pb+Pb ($\sqrt{s_{NN}}= 2.76$ TeV) collisions. From variations of the MC-Glauber and MC-KLN model parameters,
a systematic uncertainty of 2-3\% was obtained for $\bar{R}$ and $\varepsilon$ (respectively) .  
%
\begin{figure*}
\includegraphics[width=1.0\linewidth]{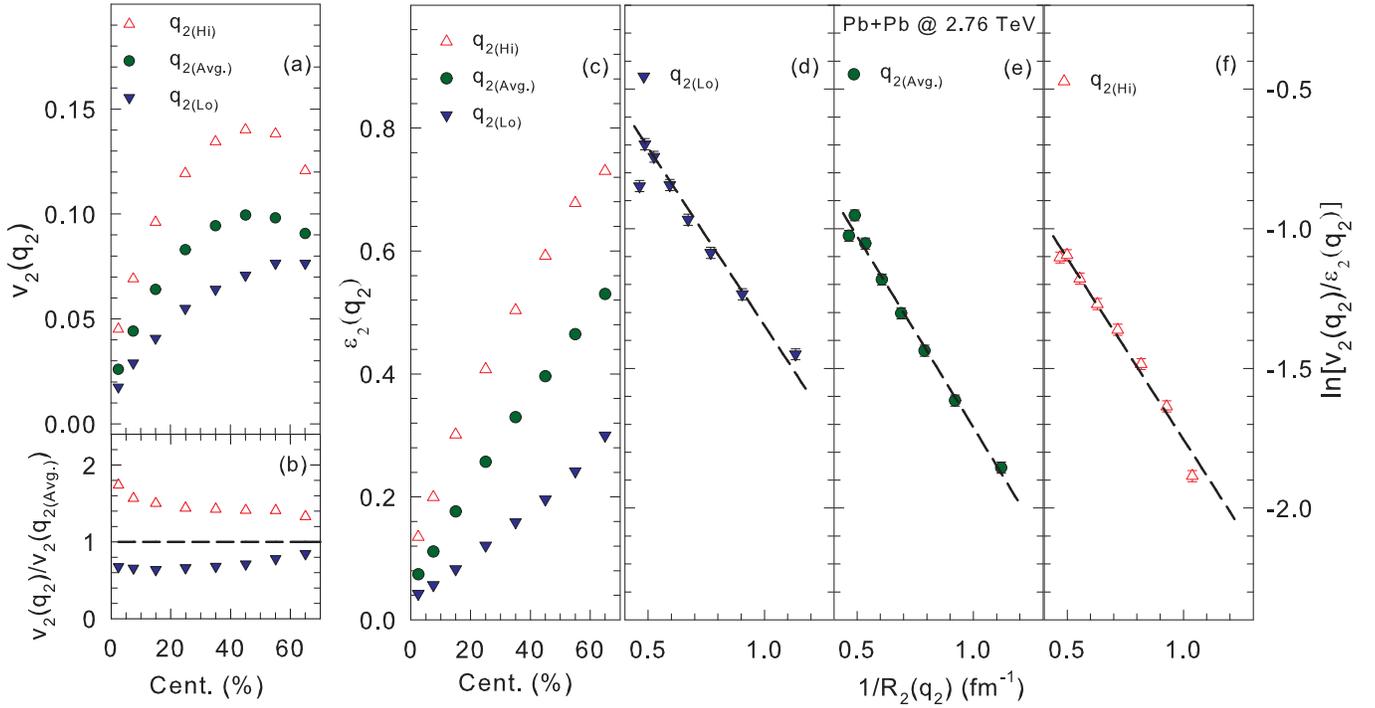}
\caption{(a) (Color online) Centrality dependence of $v_2(q_{2\text{(Lo)}})$, $v_2(q_{2\text{(Avg.)}})$ 
and $v_2(q_{2\text{(Hi)}})$ \cite{Dobrin:2012zx} for $0 <\text{ cent }< 70$\% for Pb+Pb collisions 
at  $\snn =2.76$~TeV. (b) Centrality dependence of the ratios $v_2(q_{2\text{(Lo)}})/v_2(q_{2\text{(Avg.)}})$ and
$v_2(q_{2\text{(Hi)}})/v_2(q_{2\text{(Avg.)}})$. (c) Centrality dependence of $\varepsilon_2(q_{2\text{(Lo)}})$, 
$\varepsilon_2(q_{2\text{(Avg.)}})$ and $\varepsilon_2(q_{2\text{(Hi)}})$, evaluated with the MC-Glauber model. 
(d) $\ln[v_2(q_2)/\varepsilon_2(q_2)]$ vs. $1/\bar{R}_2(q_{2})$ for $q_{2(\text{Lo})}$. (e) same as 
(d) but for $q_{2(\text{Avg.})}$. (f) same as (d) but for $q_{2(\text{Hi})}$. 
}
\label{fig:2}
\end{figure*}
%

Figure \ref{fig:1}(a) shows a representative $q_2$ distribution for 20-25\% central MC-Glauber events
for Pb+Pb collisions. The relatively broad distribution reflects the effects of sizable event-by-event 
fluctuations convoluted with statistical fluctuations due to finite particle number. 
Qualitatively similar distributions were obtained for other centralities and for other harmonics. 
These $q_n$ distributions were partitioned into the 5\% and 10\% increments 
$q_{n\text{f}}$ [from the lowest to the highest values] and used for further detailed 
selections on the event shape. 

The effectiveness of such selections is illustrated in Fig.~\ref{fig:1}(b),
which shows a strong correlation between $\varepsilon_2$  and $q_2$ for 
20-25\% central Pb+Pb events. Similar trends were obtained for other centrality 
cuts and for other harmonics. Figs.~\ref{fig:1}(c) and (d) show the dependence 
of $\varepsilon_{2}$ and $\varepsilon_{3}$ on $q_{2\text{f}}$ for two centrality 
selections as indicated. For central collisions (0-5\%), $\varepsilon_{2}(q_{2\text{f}})$
and $\varepsilon_{3}(q_{2\text{f}})$ both show an increase with $q_{2\text{f}}$, 
albeit with a much stronger dependence for $\varepsilon_{2}(q_{2\text{f}})$. 
This increase is expected to lead to 
a corresponding increase of $v_{2}(q_{2\text{f}})$ and $v_{3}(q_{2\text{f}})$ with $q_{2\text{f}}$.

Fig.~\ref{fig:1}(d) indicates a similar increase of $\varepsilon_{2}(q_{2\text{f}})$ with $q_{2\text{f}}$
for 20-25\% central collisions. However, $\varepsilon_{3}(q_{2\text{f}})$ indicates a decrease with 
$q_{2\text{f}}$, suggesting that a characteristic inversion of the dependence of $v_3(q_{2})$ 
is to be expected as a signature in future $v_3(q_{2})$ measurements for central and mid-central collisions. 

Figure~\ref{fig:2}(a) shows the centrality dependence for one set of the shape-engineered 
measurements of $v_2(q_{2\text{(Lo)}},\cent)$, 
$v_2(q_{2\text{(Avg.)}},\cent)$ and $v_2(q_{2\text{(Hi)}},\cent)$ reported in Ref.~\cite{Dobrin:2012zx}.
They show that this event-shape selection leads to lower (higher) values of 
$v_2(q_2,\cent)$ for $q_2$ values lower (higher) than $q_{2\text{(Avg.)}}$. They also show that such
selections can lead to a sizable difference (more than a factor of two) between $v_2(q_{2\text{(Hi)}},\cent)$ 
and $v_2(q_{2\text{(Lo)}},\cent)$, as illustrated in Fig.~~\ref{fig:2}(b). Strikingly similar 
differences can be observed in Fig.~\ref{fig:2}(c) for the MC-Glauber results shown for $\varepsilon_2(q_{2\text{(Lo)}},\cent)$, 
$\varepsilon_2(q_{2\text{(Avg.)}},\cent)$ and $\varepsilon_2(q_{2\text{(Hi)}},\cent)$. They suggest 
that differences in the measured magnitudes for $v_2(q_{2\text{(Lo)}},\cent)$, 
$v_2(q_{2\text{(Avg.)}},\cent)$ and $v_2(q_{2\text{(Hi)}},\cent)$, are driven by 
the corresponding differences in the calculated magnitudes for $\varepsilon_2(q_{2\text{(Lo)}},\cent)$, 
$\varepsilon_2(q_{2\text{(Avg.)}},\cent)$ and $\varepsilon_2(q_{2\text{(Hi)}},\cent)$.

The shape-selected measurements in Fig.~\ref{fig:2}(a) for $v_2(q_{2\text{(Lo)}},\cent)$, $v_2(q_{2\text{(Avg.)}},\cent)$ 
and $v_2(q_{2\text{(Hi)}},\cent)$ all show an increase from central to mid-central collisions, 
as would be expected from an increase in $\varepsilon_2(q_{2\text{(Lo)}},\cent)$, 
$\varepsilon_2(q_{2\text{(Avg.)}},\cent)$ and $\varepsilon_2(q_{2\text{(Hi)}},\cent)$
over the same centrality range [cf. Fig.~\ref{fig:2}(c)]. For $\cent \agt 45\%$ however, the decreasing 
trends for $v_2(q_{2\text{(Lo)}},\cent)$, $v_2(q_{2\text{(Avg.)}},\cent)$ and $v_2(q_{2\text{(Hi)}},\cent)$ 
contrasts with the increasing trends for $\varepsilon_2(q_{2\text{(Lo)}},\cent)$, 
$\varepsilon_2(q_{2\text{(Avg.)}},\cent)$ and $\varepsilon_2(q_{2\text{(Hi)}},\cent)$, 
suggesting that the viscous effects due to the smaller systems produced in peripheral collisions, serve to 
suppress $v_2(q_{2\text{(Lo)}},\cent)$, $v_2(q_{2\text{(Avg.)}},\cent)$ and $v_2(q_{2\text{(Hi)}},\cent)$. 
This is confirmed by the symbols and dashed 
curves in Figs.~\ref{fig:2}(d)~-~(f) which validates the expected linear dependence 
of $\ln[v_2(q_2)/\varepsilon_2(q_2)]$ on $1/\bar{R}_2(q_{2})$ (cf. Eq.~\ref{eq:2}) 
for the data shown in Fig.~\ref{fig:2}(a). The dashed curves, which indicate a similar 
slope value ($\beta'' \sim 1.3 \pm 0.07$) for each of the scaling curves in 
Figs.~\ref{fig:2}(d)~-~(f), provide an invaluable model constraint for the
event-by-event fluctuations in the initial-state density distribution, as well as 
for robust estimates of $\eta/s$.
%
\begin{figure}
	\includegraphics[width=1.0\linewidth]{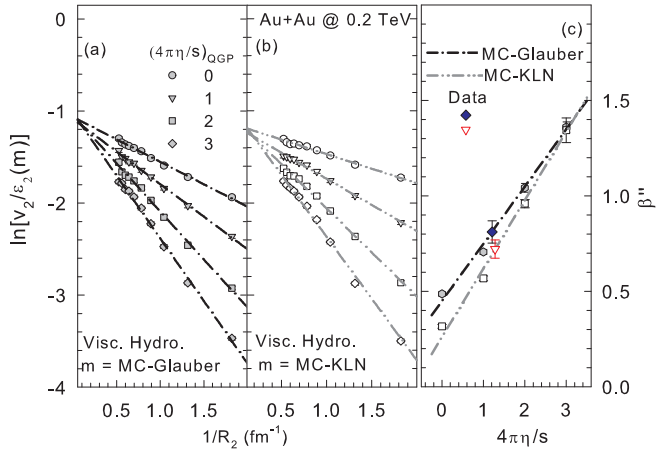}
  \caption{(Color online) $\ln[v_2/\varepsilon_2]$ vs. $1/\bar{R}_2$ for 
viscous hydrodynamical calculations \cite{Song:2011hk} for Au+Au collisions 
at $\snn =0.2$~TeV with 
(a) MC-Glauber initial-state geometries and (b) MC-KLN initial-state geometries;
the dashed-dot and the dotted-dashed curves represent linear fits. Results
are shown for several values of $4\pi\eta/s$ as indicated. (c) Calibration curve 
for $\beta''$ vs. $4\pi\eta/s$; the $\beta''$ values are obtained from the slopes of 
the curves shown in (a) and (b). 	The indicated data points are obtained from 
a linear fit to $\ln[v_2/\varepsilon_2]$ vs. $1/\bar{R}_2$ for the STAR Au+Au data  
at $\snn =0.2$~TeV \cite{Adams:2004bi,Aamodt:2010pa}
}
\label{fig:3}
\end{figure}
%

\begin{figure}
	\includegraphics[width=1.0\linewidth]{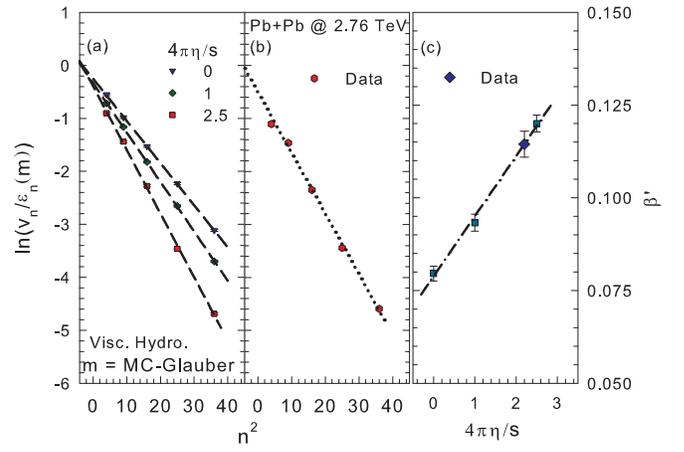}
  \caption{(Color online) (a) $\ln(v_n/\varepsilon_n)$ vs. $n^2$ from viscous 
hydrodynamical calculations \cite{cms_ulc_note} for three values of specific 
shear viscosity as indicated. (b) $\ln(v_n/\varepsilon_n)$ vs. $n^2$ for Pb+Pb data.
The $p_T$-integrated $v_n$ results in (a) and (b) are for 0.2\% central Pb+Pb collisions
at $\sqrt{s_{NN}}= 2.76$ TeV \cite{cms_ulc_note}; the curves are linear fits. 
(c) Calibration curve for $\beta'$ vs. $4\pi\eta/s$; the $\beta'$ values are obtained 
from the slopes of the curves shown in (a). 	The indicated data point is obtained from 
a linear fit to the scaled data shown in (b).
}
  \label{fig:4}
\end{figure}

The acoustic scaling patterns summarized in  Eq.~\ref{eq:2} are also exhibited in 
the results of $q_n$-averaged viscous hydrodynamical 
calculations \cite{Song:2011hk,*cms_ulc_note} as demonstrated  
in Figs.~\ref{fig:3}(a) and (b) and Fig.~\ref{fig:4}(a). The scaled results, which are 
shown for several values of $4\pi\eta/s$ in each case, exhibit the expected linear 
dependence of $\ln(v_n/\varepsilon_n)$ on $1/\bar{R}$ for both MC-Glauber (Figs.~\ref{fig:3}(a)) 
and MC-KLN (Figs.~\ref{fig:3}(b)) initial conditions, as well as the  expected linear 
dependence of $\ln(v_n/\varepsilon_n)$ on $n^2$ (Fig.~\ref{fig:4}(a)). 
They also give a clear indication that the slopes of these curves 
are sensitive to the magnitude of $4\pi\eta/s$. Therefore, we use them to 
calibrate $\beta''$ and $\beta'$ to obtain estimates for $(4\pi\eta/s)_\mathrm{QGP}$ 
for the plasma produced in RHIC and LHC collisions.

Figure~\ref{fig:3}(c) shows the calibration curves for $\beta''$ vs. $4\pi\eta/s$, obtained 
from the viscous hydrodynamical calculations shown in Figs.~\ref{fig:3}(a) and (b).
The filled circles and the associated dot-dashed curve, represent the slope parameters ($\beta''$) 
obtained from linear fits to the viscous hydrodynamical results for MC-Glauber initial 
conditions shown in Fig.~\ref{fig:3}(a). 
The open squares and the associated dot-dot-dashed curve, represent the slope parameters obtained 
from linear fits to the viscous hydrodynamical results for MC-KLN initial conditions 
shown in Fig.~\ref{fig:3}(b).
The STAR $v_2(\cent)$ data for Au+Au collisions, also show the expected linear dependence 
of $\ln(v_2/\varepsilon_2)$ on $1/\bar{R_2}$ for $\varepsilon_2$ and $\bar{R_2}$ values obtained 
from the MC-Glauber and MC-KLN models respectively. The filled diamond and the open triangle 
in Fig.~\ref{fig:3}(c), represent the slopes extracted from the respective scaling 
plots for MC-Glauber and MC-KLN initial conditions; 
a comparison to the respective calibration curves in Fig.~\ref{fig:3}(c), gives 
the estimate $\la 4\pi\eta/s\ra_\mathrm{QGP}\sim 1.3 \pm 0.2$ for the plasma created 
in RHIC collisions. Here, it is noteworthy that our extraction procedure leads to 
an  estimate which is basically insensitive to the choice of the MC-Glauber or MC-KLN 
initial-state geometry. 
%
%

The solid squares and the associated dashed-dot curve in Fig.~\ref{fig:4}(c), represent the 
calibration curve for $\beta'$ vs. $4\pi\eta/s$, obtained from the linear fits (dashed curves) 
to the viscous hydrodynamical calculations shown in Fig.~\ref{fig:4}(a). 
Fig.~\ref{fig:4}(b) shows the expected linear dependence of $\ln(v_n/\epsilon_n)$ on $n^2$ 
for CMS Pb+Pb data  \cite{cms_ulc_note} scaled with same $\varepsilon_n$ values employed 
in Fig.~\ref{fig:4}(a). The slope extracted from Fig.~\ref{fig:4}(b) is indicated 
by the solid blue diamond shown in Fig.~\ref{fig:4}(c); a comparison with the 
the calibration curve gives the the estimate 
$\la 4\pi\eta/s\ra_\mathrm{QGP}\sim 2.2 \pm 0.2$ for the plasma created 
in LHC collisions.

The $\la 4\pi\eta/s\ra_\mathrm{QGP}$ estimates for the plasma produced in RHIC and LHC 
collisions
are in reasonable agreement with recent $\left< \eta/s \right>$
estimates \cite{Lacey:2011ug,Schenke:2011tv,*Schenke:2011bn,Gale:2012in,Qiu:2011hf}.
While further calculations will undoubtedly be required to reduce possible
model-driven calibration uncertainties, our method benefits from the implicit  
constraint for the event-by-event fluctuations in the initial-state density distribution,
as well as its lack of sensitivity to the initial-state models employed in our 
analysis.


In summary, we have presented a detailed phenomenological exploration of a new constraint 
for initial-state fluctuations, via scaling studies of $v_2$ measurements obtained for 
shape-engineered events. We find acoustic scaling patterns for shape-selected events
(via $q_{2\text{(Lo)}}$, $q_{2\text{(Avg.)}}$ and $q_{2\text{(Hi)}}$) which 
provide robust constraints for the event-by-event fluctuations in the initial-state 
density distribution, as well as methodology with two consistent paths 
for estimating $(\eta/s)_\mathrm{QGP}$ of the QGP
produced in Au+Au and Pb+Pb collisions at RHIC and the LHC. A calibration of the method 
with $q_2$-averaged viscous hydrodynamical model calculations, gives estimates for 
$(4\pi\eta/s)_\mathrm{QGP}$ of $1.3 \pm 0.2$ and $2.2 \pm 0.2$,
for the plasma produced in Au+Au ($\sqrt{s_{NN}}= 0.2$ TeV) 
and Pb+Pb ($\sqrt{s_{NN}}= 2.76$ TeV) collisions (respectively). These values are 
insensitive to the initial-state geometry models employed.

{\bf Acknowledgments}
This research is supported by the US DOE under contract DE-FG02-87ER40331.A008. 
 




%
\bibliography{evnt_eng} 
\end{document}